\documentclass[11pt]{article}

\usepackage{urcsreport,urcsbiblio,urcsversions,url}
\usepackage{graphicx}

\begin{document}

\title{Analysis and Counterexamples Regarding Yatsenko's Polynomial-Time Algorithm for Solving the Traveling Salesman Problem}
\author{Christopher Clingerman \and Jeremiah Hemphill \and Corey Proscia}
\techrepname{Technical Report}
\techrepnum{}
\maketitle

\begin{abstract}

Yatsenko gives a polynomial-time algorithm for solving the traveling salesman problem. We examine the correctness of the algorithm and its construction. We also comment on Yatsenko's evaluation of the algorithm.

\end{abstract}

\section{Introduction}
\label{introduction}

In the study of computer science, one of the most interesting and difficult questions is whether the set of $P$ problems equals the set of $NP$ problems. $P$ is the class of languages that are decidable in polynomial time on a deterministic Turing machine \cite{sipser}. $NP$ is the class of languages that have polynomial-time verifiers. Currently it is unknown whether $P = NP$, but if those two complexity classes are indeed equivalent, it means that all problems in $NP$ can be solved in polynomial time. However, if it turns out that $P \neq NP$, every one of the hundreds of important, natural $NP$-complete problems is not solvable in polynomial time.  The implications of $P$ and $NP$ equivalence or inequality are incredibly important to the computer science community and to related fields.  

One particularly important problem that is currently known to be in $NP$ is the traveling salesman problem. This problem is also $NP$-complete because it is in $NP$ and every problem in $NP$ is reducible to it in polynomial time \cite{sipser}, so if the traveling salesman problem can be solved in polynomial time, then every problem in $NP$ can be solved in polynomial time and $P = NP$. The Traveling Salesman Problem is stated thusly: a traveling salesman has a table of distances between $N$ cities.  He wishes to travel to each of these cities in turn and try to sell his wares.  He also wants to minimize the distance he travels in order to save time and travel expenses.  However, he can only visit each city once, and he must finish at the same city from which he started. A brute force solution that examines every possible route would examine all $N!$ routes, which is impractical for even moderately large values of $N$. As one can see, the complexity of this problem seemingly quickly increases with the number of cities. In his paper \cite{yatsenko2007fem}, Vadim Yatsenko claims to have found an algorithm that both solves the traveling salesman problem and runs in polynomial time. However, Yatsenko's algorithm on some inputs produces incorrect results, and even Yatsenko's paper (on its second page) concedes that it is less reliable for larger values of $N$.

\section{Description of the Algorithm}
\label{description}

Yatsenko gives an algorithm for solving the traveling salesman problem \cite{yatsenko2007fem}. The algorithm is broken into three steps. The first two steps describe how to start the algorithm, and the third step is applied iteratively. The algorithm starts by creating a route with three points, which forms a triangle. This triangle is computed by connecting the two points that are the farthest apart. A third point is added so that the route connecting all three points, or the sum of the distances between the three pairs, is the largest possible.

After the triangle is constructed, points are added one at a time. For each edge on the route, a third point is selected whose addition to the route, replacing the edge, would change the length of the route by the smallest amount. This change of length is referred to as the disturbance and is the sum of the two added edges minus the removed edge. Then, out of all of the thus chosen third points (one per edge), the one whose addition creates the greatest disturbance is added to the route. This process is repeated, adding points to the route one at a time, until all points are added. At each step, the point with the maximum of the minimum disturbances is added.

\section{Correctness of the Algorithm}
\label{analysis}

\subsection{The Max-min Algorithm}

Yatsenko claims that the inverse of a cutting procedure gives an adding procedure that gives the optimal solution. The adding procedure that Yatsenko describes is not the inverse of the cutting procedure, and it does not reconstruct the optimal solution. The cutting procedure begins with an optimal route, the solution to a traveling salesman problem. At each step of the cutting procedure, Yatsenko removes the point that gives the least disturbance to the length of the route. When the point is removed, the two points adjacent to it on the route are connected, and the magnitude of the length of the added path minus the two removed is the disturbance. The magnitude of the disturbances increases as each vertex is cut from the route.

\begin{figure}
  \begin{center}
	\includegraphics[height=2in,width=2in]{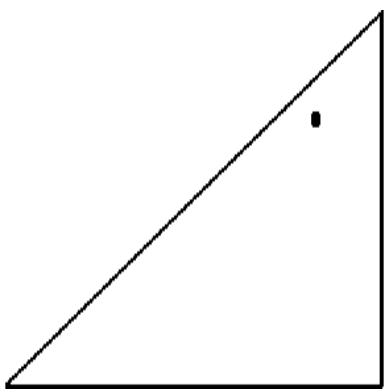}
     \caption{initial setup for max-min}
     \label{f:max-min-problem}
  \end{center}
\end{figure}

\begin{figure}
  \begin{center}
   	\includegraphics[height=2in,width=2in]{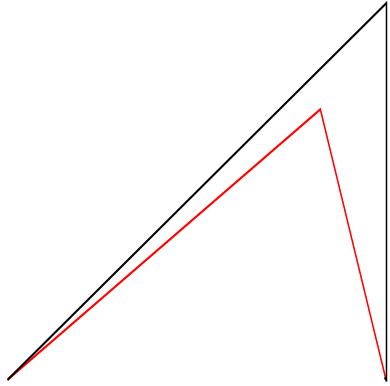}
    \caption{max-min incorrect solution}
    \label{f:max-min-incorrect}
  \end{center}
\end{figure}

\begin{figure}
  \begin{center}
   	\includegraphics[height=2in,width=2in]{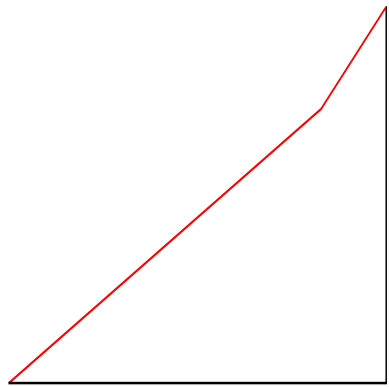}
    \caption{max-min correct solution}
    \label{f:max-min-correct}
  \end{center}
\end{figure}

In the final step of the adding procedure, there is only one choice of which point to use for each edge in the current route. Of the ways to replace these edges with two edges to the point, according to the algorithm, the one that gives the maximum disturbance should be chosen. Consider Figure~\ref{f:max-min-problem}, which gives the initial triangle and a fourth point of a graph. This example is in Euclidean space, although Yatsenko claims that his algorithm works in the general case. The initial triangle was constructed by connecting points A and B, which are the farthest apart, and adding point C to increase the route's length by the largest amount. For each edge, the only remaining point is selected, and the edge whose replacement gives the largest displacement is removed. This gives the route shown in Figure~\ref{f:max-min-incorrect}, while the optimal route is shown in Figure~\ref{f:max-min-correct}. This example shows that the adding procedure does not give the optimal route, and it is not the inverse of the cutting procedure.

\subsection{The Min-min Algorithm}

An alternative adding procedure is to consider changing the route by the least amount possible at each step, starting with the same initial triangle, so we also consider the min-min case as both an inverse to the cutting procedure and with a counter-example of correctness.

At each step, the point with the minimum disturbance is added. For example, a node that gives the minimum disturbance is added, followed by adding another node with minimum disturbance. If a node is selected to be cut, it would be the first node added, not the most recent, assuming that the two nodes are not adjacent. If the node with the minimum disturbance is added and then the node with the minimum disturbance is subtracted, they are not always the same, so the cutting procedure and this adding procedure do not remove and add points in the same order. Since adding a point and then cutting a point does not always add and cut the same point, the min-min adding procedure and the cutting procedure do not correspond.

\begin{figure}
  \begin{center}
   	\includegraphics[height=2.5in,width=2.5in]{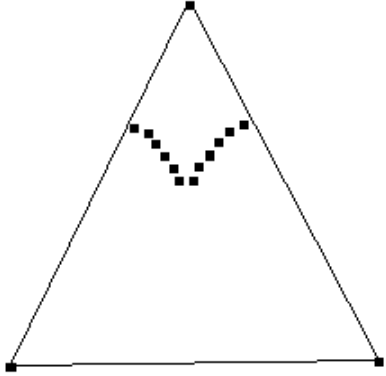}
    \caption{initial setup for min-min}
    \label{f:min-min-setup}
  \end{center}
\end{figure}

\begin{figure}
  \begin{center}
   	\includegraphics[height=2.5in,width=2.5in]{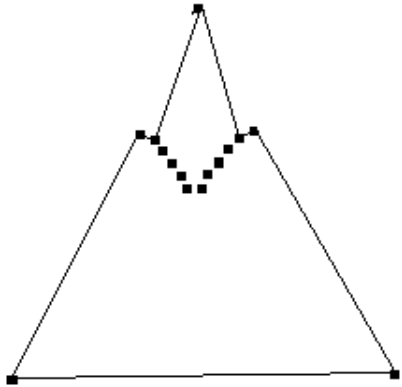}
    \caption{points are processed inward from the outside}
    \label{f:min-min-process}
  \end{center}
\end{figure}

\begin{figure}
  \begin{center}
   	\includegraphics[height=2.5in,width=2.5in]{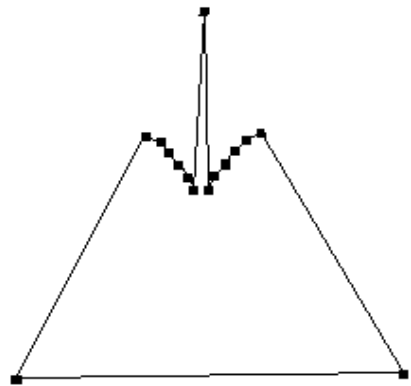}
    \caption{actual route with the min-min algorithm}
    \label{f:min-min-incorrect}
  \end{center}
\end{figure}

\begin{figure}
  \begin{center}
   	\includegraphics[height=2.5in,width=2.5in]{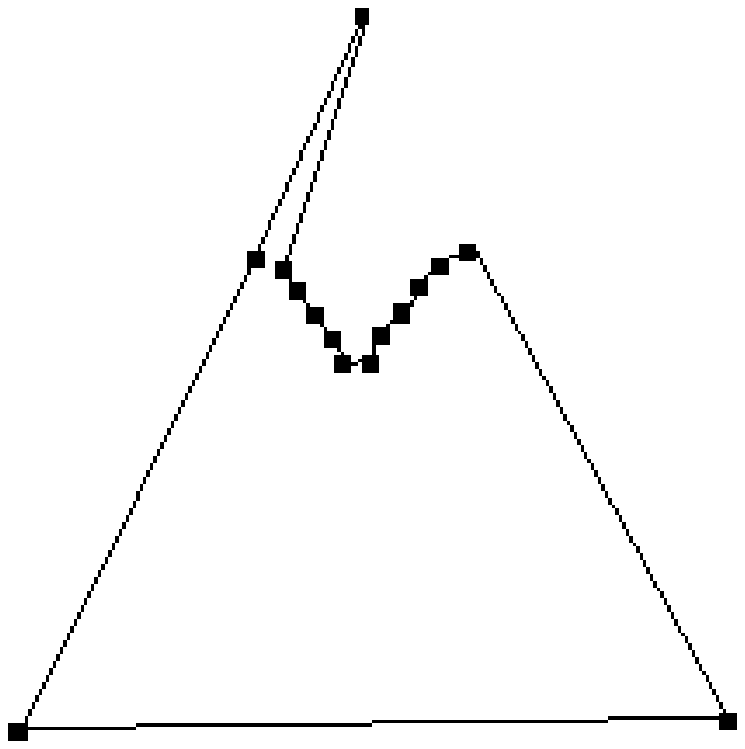}
    \caption{optimal route}
    \label{f:min-min-correct}
  \end{center}
\end{figure}

Specific graph constructions can also lead to counter-examples for this algorithm.  Simply checking for the absence of intersections is not sufficient to show that the route is optimal. Consider the set of points in Figure~\ref{f:min-min-setup} with the initial three routes already determined. The three outer points form the initial triangle and the rest of the points fall inside of it. The distances between the points can be arranged so that the algorithm will process the inner points in sequence, moving from the outside towards the center as in Figure~\ref{f:min-min-process}. Once all the points have been included, the route is shown in Figure~\ref{f:min-min-incorrect}, which is not the optimal route. The optimal route should connect the top point to the other points with the shortest distance possible. Since the distance from the top to the side is smaller than the distance from the top to the middle, the optimal route is shown in Figure~\ref{f:min-min-correct}.

\section{Exponential Subcases}
\label{exponential}

As Yatsenko mentions, situations can occur when multiple nodes satisfy the condition to be added. For all but the initial route of two points, this set of solutions can be denoted
\[\max_{r_{k} \in \overline{R^{i}}} \min_{p_{j} \in P^{N} - \overline{R^{i}}} \Delta(r_{k}, p_{j}) \rightarrow ({r_{k}, p_{l1}}), ({r_{m1},p_{l1}})...\]
It may be necessary, if one assumes that tied cases are viable and must be brute-force explored, to search each solution along all possible variations of $r_{j}^{i+1}$ intermediate sub-routes, where $j$ is the number of possible solutions. To solve this problem, the algorithm must essentially run itself once for each possible solution from that point forward. In addition, subsequent steps may also have multiple subproblems, each of which must be examined separately. When this situation occurs the run-time of the algorithm can no longer be considered $O(N^{3})$.

\begin{figure}
  \begin{center}
	\includegraphics[height=2.5in,width=2.5in]{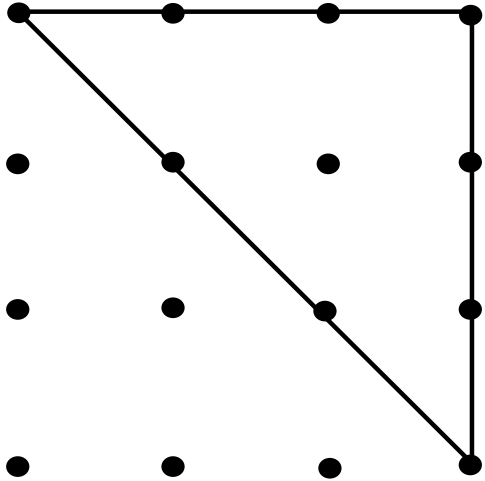}
     \caption{example situation causing exponential subcases}
     \label{f:exponential}
  \end{center}
\end{figure}

One degenerate situation is a square grid of equally spaced nodes in patterns of squares (see Figure~\ref{f:exponential}.) In this case, nearly every decision by the algorithm leads to multiple subproblems.

For the initial $R^{2}$ route, there are two different choices, the major and minor diagonals. The $R^{2}$ route calculates the maximum distance instead of the minimum and the solutions are denoted, \[\max_{p_{i}, p_{j} \in P^{N}} d(p_{i}, p_{j}) \rightarrow ({p_{k1}, p_{l1}}), ({p_{k2},p_{l2}})...\]  Similarly, for the third node, there are two different solutions.  Based on which diagonal is chosen, the two possible solutions are the corners opposite the diagonal.  After selecting two nodes, the algorithm now must calculate four iterations simultaneously.

Once the initial maximum size triangle is created, the algorithm searches for the minimum increase in route distance.  In this case, there are $3(N-2)$ points that fall on the maximum triangle.  Although an intelligent algorithm could ignore these points and add them automatically, this is not specifically handled in the paper.  The algorithm must iterate over every node to consider all the solutions.  In addition, after the first set of solutions is handled, the remaining $3(N-2)-1$ nodes must be considered for each iteration.  For a 100 node graph, there are an additional $(3\times(10-2))!$ iterations of the algorithm, one for each permutation.

After these zero-length increases of disturbance, the smallest increase in total route length comes from changing a diagonal into two horizontal or vertical lines. If the distance between nodes is $d$, this procedure adds $d\times\sqrt{2} - d$ distance to the total route.  A solution set of $2(N-2)$ nodes adjacent to the major or minor diagonal must be considered.  If the algorithm iterates on a node on the corner of the grid, $2(N-2)-1$ nodes must still be considered.  On any other node in the solution set, $2(N-2)-2$ nodes will be in the solution set in the next step.  For a 100 node graph, this is an increase of $9!$ iterations of the algorithm.

Once all the diagonals have been removed, a two step procedure of adding a diagonal and removing it can be used to fill in the rest of the graph.  The solution set consists of each node adjacent to the current route.  Each node has an average of two ways to connect it to the graph using a single diagonal and a single horizontal or vertical addition to the route.  Once this change is made, the minimal route increase is to add the node that removes the diagonal.  This process is repeated to add the rest of the nodes to the route. This will add a significant number of iterations to the algorithm.

Almost every decision in this graph requires simulating additional instances of the route.  Due to this, the algorithm no longer operates in polynomial time. A pruning method may be possible to avoid considering overlapping subproblems, but this is not specified in the algorithm. A graph with a grid or subsection of a grid of equally spaced nodes will always have this problem.   

\section{Comments on Yatsenko's Evaluation}
\label{visual}

In his paper, Yatsenko informs the reader that in order to verify his algorithm's success in finding the solution to a data set, he uses what he calls ``visual inspection."  Also provided in the paper are several images of data sets with superimposed solutions for $N =$ 500, 1000, and 2000.  Upon analysis of these images, however, one finds that a visual inspection approach to verify a solution is not accurate.  This may be a good tool for very small values of $N$, but for values of 500 and above, it is not an accurate method of determining the correctness of a solution.

An alternative to visual inspection would be to use instances from standard data sets that have known solutions \cite{vlsi}. After running the algorithm on this data set, the total distance of the route the algorithm gives could be compared with the published result. Many heuristics also exist for the traveling salesman problem, which may provide a less thorough method of verifying a result for an unpublished instance.

Yatsenko also admits that his algorithm forms ``loops" on a growing percentage of solutions over random data sets as $N$ increases. We interpret loops to mean the route intersects with itself. These solutions are not optimal because unwinding the intersection always gives a shorter route. For problems with 2000 points, Yatsenko states these intersections form 4 out of 5 times. Even if Yatsenko's algorithm succeeds in operating in polynomial time, as the value of $N$ increases his algorithm becomes more unstable in that it produces incorrect solutions with greater frequency. While the algorithm may work for some $N$, it is not a valid general solution for the traveling salesman problem.

\section{Conclusion}
\label{conclusion}

Yatsenko's solution to the traveling salesman problem does not always produce optimal results, so it cannot be clearly considered an exact (in the sense of optimal) solution. It may be considered a heuristic, but he does not evaluate it as one. Yatsenko's own evaluation of the algorithm reveals some cases where it gives incorrect solutions even though the evaluation does not detect all errors. There are cases where the subproblems formed by equidistant vertices overlap, but the algorithm does not specify how to eliminate them, so there may be an exponential number of subproblems, and the running time of the algorithm would not be polynomial in the worst case. 

\section{Acknowledgments}
\label{thanks}

We thank Lane Hemaspaandra and Saurabh Deshpande, the professor and teaching assistant for the course in which we started this note, for helpful feedback on earlier drafts and support in editing.  

\bibliographystyle{alpha}
\bibliography{sources}

\end{document}